\documentclass[aps,a4paper,twocolumn,showpacs,hyphens]{revtex4}
\usepackage{epsfig}
\usepackage{graphicx}
\usepackage{amsmath,amssymb,color}
\usepackage[english]{babel}																					
\usepackage[colorlinks=true, allcolors=blue]{hyperref}
\usepackage{comment}
\usepackage{adjustbox}
\usepackage{orcidlink}

\parskip=\medskipamount

\newcommand{\eq}[1]{(\ref{#1})}
\newcommand{\fig}[1]{Fig.~\ref{#1}}

\newcommand{\be}{\begin{equation}}
\newcommand{\ee}{\end{equation}}

\newcommand{\barr}{\begin{array}}
\newcommand{\earr}{\end{array}}

\newcommand{\beqn}{\begin{eqnarray}}
\newcommand{\eeqn}{\end{eqnarray}}

\newcommand{\bs}{\begin{subequations}}
\newcommand{\es}{\end{subequations}}

\newcommand{\bw}{\begin{widetext}}
\newcommand{\ew}{\end{widetext}}

\newcommand{\st}[1]{{\scriptscriptstyle #1}} 
\newcommand{\eff}{\mathrm{eff}} 

\begin{document}



\title{Threshold model of language competition including the bilingual state}

\author{M.V.~Tamm$^1$\orcidlink{0000-0003-3168-1307}, E.~Heinsalu$^2$, S.~Scialla$^{2,3}$\orcidlink{0000-0003-1582-8743}, M.~Patriarca$^{2,4}$\orcidlink{0000-0001-6743-2914}}

\affiliation{$^1$School of Digital Technologies, Tallinn University, Tallinn, Estonia \\
$^2$National Institute of Chemical Physics and Biophysics, Tallinn, Estonia
\\
$^3$Department of Science and Technology for Sustainable Development and One Health, Università Campus Bio-Medico di Roma, Rome, Italy \\
$^4$Department of Cybernetics, Tallinn University of Technology, 19086 Tallinn, Estonia}

\date{\today}

\begin{abstract}
%
%
We propose a threshold model of language competition which includes an intermediate bilingual state. 
The model is based on the Minett-Wang model but through the introduction of thresholds in the language shift rates, it incorporates the effects of memory and learning. 
The model is piecewise-linear, allowing an exact analytical treatment.
We study the symmetric case where two competing languages are equivalent in terms of status and social pressure and provide a complete list of the various dynamical regimes. 
We also study several limiting regimes corresponding to asymmetric systems and characterize the full spectrum of possible asymptotic behaviors. 
Unlike the Minett-Wang model, which always predicts the extinction of one of the languages, the proposed new model exhibits a wide range of possible equilibrium scenarios, including equilibrium states of coexistence.  
Most commonly, in such coexistence regimes the minority language speakers are either predominantly monolingual or predominantly bilingual.

\end{abstract}


\maketitle

\section{Introduction} 

Due to the high social relevance and an increasing occurrence of multilingual societies on a global scale, mathematical modeling of language competition has become a research topic of strong interest \cite{Patriarca-2020a}. 
Since the foundational works of Baggs and Freedman \cite{Baggs-1990a,Baggs-1993a,Wyburn2008a,Wyburn2009a,Wyburn2010a}, Abrams and Strogatz (AS) \cite{Abrams-2003a}, and Minett and Wang (MW) \cite{Minett2008a}, language competition dynamics has been studied using different frameworks, from analytical models mathematically analogous to those used to describe ecological competition and chemical reactions kinetics, to heterogeneous network and many-agent models.
A limitation of the models developed in the named papers, as well as of other similar models -- both two-state models describing populations that are comprised of monolingual speakers of the two competing languages,  and three-state models, which also include bilingual speakers -- is that the form of their dynamical equations leads to the eventual extinction of one of the languages \cite{Abrams-2003a,Minett2008a,Baronchelli-2006c}, unless certain conditions are assumed \cite{Castellano-2009a,Heinsalu-2014b}.
However, this is inconsistent with the observation that two or more languages often coexist in the same society for centuries.

Recently, we proposed a language competition model that is based on the AS model, but where the language shift rates are redefined so that they allow to address the importance of the language learning processes as well as the population heterogeneity in language competition \cite{Tamm-2025}. 
On a coarse-grained timescale, the effects of memory and learning in the dynamics of language shift can be expressed as thresholds on the speakers’ fractions of the competing languages, i.e., the power-law form of the language shift rates in the AS model is replaced with a step function, when all individuals in the population are assumed to be identical, or with a sigmoid function, when population heterogeneity is taken into account. 
Importantly, besides the extinction of one of the two languages, the threshold model of Ref.~\cite{Tamm-2025} predicts two other long-term regimes, corresponding to language coexistence.

The effects of memory and language learning process, as well as attrition, were studied in Ref.~\cite{Scialla-2023a} via agent-based simulations, by incorporating them into a three-state model and analyzing how they influence the dynamical evolution of the system. 
The model of Ref.~\cite{Scialla-2023a} is built on two assumptions.
First, concerning the learning process, memory was taken into account, assuming that a monolingual agent can acquire the other language and become bilingual only through repeated interactions with monolingual speakers of the other language, i.e., a monolingual agent has to use the new language a certain number of times within a certain time interval \cite{Ebbinghaus-1885a,Murre-2015a,Sebastián-Gallés_2015}. 
Furthermore, as far as the attrition process is concerned, in order to maintain a language, a bilingual speaker has to use it with a certain minimum frequency \cite{Schmid-2019a,Higby_Obler_2015}.

During the last two decades there has been a significant interest in bilingual language learning and processing and as a result a number of computational models have been developed that account for bilingualism \cite{Li-2023}.
The goal of the present paper is to generalize the threshold model of language competition introduced in Ref.~\cite{Tamm-2025} to the case of a three-state MW-type model, which explicitly takes into account the presence of bilinguals. 
This implies replacing the learning rates with threshold-based rates, in a way similar to Ref.~\cite{Tamm-2025}, and an analogous modification of the attrition rates, as we will discuss below. 
Even if the introduction of the intermediate bilingual state leads to additional complexity, the equations of the model are still piecewise-linear, which allows for an analytical study of the dynamical trajectories.
As a result, a rich pattern of possible evolution outcomes is observed, producing a variety of regimes similar to those observed in real bilingual societies.

The paper is organized as follows. In Sec.~\ref{sec:Models} we discuss the existing three-state models of language competition and propose the new threshold-based model.
In Sec.~\ref{sec:MW-model} we briefly recall the MW model. 
In Sec.~\ref{sec:bilinguals} we discuss how the bilinguals can influence learning and attrition processes. 
Then, in Sec.~\ref{sec:TMW-model}, we introduce thresholds in the language shift rates and formulate the model. 
The solution of the model is presented in Sec.~\ref{sec:results}. We start in Sec.~\ref{sec:pure} with the discussion of ``pure'' regimes, which we use further on as building blocks for the solution. Then, in Sec.~\ref{sec:example} we solve the model for one exemplary set of parameters. In Sec.~\ref{sec:limiting} we discuss limiting cases in which language attrition is either always present or always absent, independently of the other parameters. In Sec.~\ref{sec:symmetric} we give a full solution  in case when there is a complete symmetry between languages in terms of their complexity and prestige. Finally, in Sec.~\ref{sec:conclusion} we summarize our results, discuss possible comparison with observational data and make some concluding remarks.

\section{Three-state models with bilinguals} \label{sec:Models}

%
%

Models of language dynamics with bilinguals, such as the MW model \cite{Minett2008a}, are three-state models, meaning that there are three types of speakers: from the total population a fraction $x$ are monolinguals of language X, a fraction $y$ are monolinguals of language Y, and a fraction $z$ are bilinguals, denoted in the following by Z, who speak both languages X and Y. 
Here we assume that the total population is constant at any time, $x + y + z = 1$.
This assumption means that the model validity is limited to situations of homogeneous population growth, i.e., the reproduction and access to resources are similar in the interacting linguistic groups (see Ref.~\cite{Heinsalu-2014b} for a discussion of this point).

\begin{figure*}[t!]
\includegraphics[width=17cm]{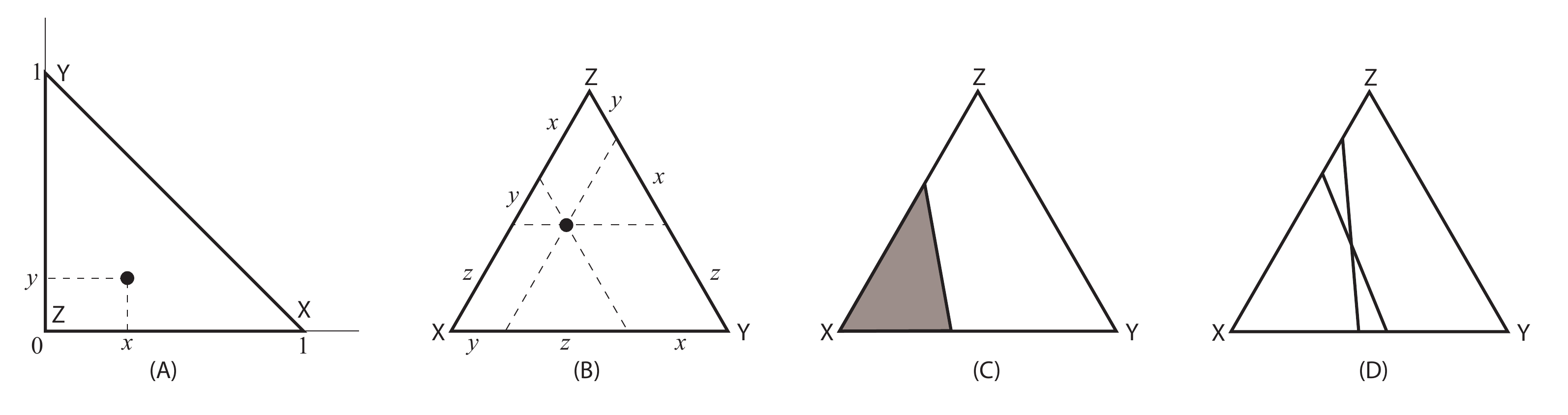}
\caption{
Panels (A) and (B) represent the state of a three-state system in Cartesian coordinates and by a ternary diagram, respectively. 
Panel (C): learning and attrition thresholds \eqref{Lrates-fractions-1} and \eqref{eq_newrates-attrit} correspond to straight lines, each of them splits the ternary diagram into regions where the corresponding process is on or off;
Panel (D): whenever parameters $\beta^+$ and $\beta^-$ are unequal, the learning and retention thresholds may intersect; if $\beta^- > \beta^+$ the steeper line corresponds to the retention threshold.}
\label{ternary}
\end{figure*}

There are two common ways to visualize the state of such a three-state system. 
One possibility, depicted in \fig{ternary}A, is to eliminate one of the variables, e.g., $z = 1 - x - y$, and to use the conventional Cartesian plane of the remaining coordinates, $x$ and $y$. 
Another possibility is to depict the state of a system as a point inside a unilateral triangle with sides of unit length as illustrated in \fig{ternary}B; such depiction is commonly called a ternary diagram. 
A ternary diagram is nothing but an affine transformation of the Cartesian representation.
In a ternary diagram, the corner labeled ``X'' corresponds to the pure state of monolinguals X, $x = 1$; while the opposite side YZ of the diagram corresponds to states where $x = 0$, i.e., there are no monolinguals of language X but only monolinguals of language Y and bilinguals Z, $y + z = 1$.
The same is true for corners labeled ``Y'' and ``Z'' and the opposite sides XZ and XY, respectively. 
Figure \ref{ternary}B illustrates the correspondence between the position of an arbitrary point and the corresponding values $x, y, z$. 
In what follows we use the ternary diagram representation, which preserves the symmetry between the variables.  

%

\subsection{Minett-Wang model} \label{sec:MW-model}

In the MW model, no direct transitions between states X and Y are possible, X $\not\rightleftarrows$ Y, reflecting the fact that shifting from language X  to language Y, and vice versa, can happen only by passing through the bilingual state Z. 
Thus, only the transitions X $\rightleftarrows$ Z $\rightleftarrows$ Y can take place: transitions X $\to$ Z and Y $\to$ Z correspond to language acquisition, i.e. monolinguals becoming bilingual; transitions Z $\to$ X and Z $\to$ Y correspond to language attrition, i.e., bilinguals becoming monolingual. 
The time evolution of the fractions of mono- and bilingual speakers follow the equations~\cite{Minett2008a}
\begin{equation} \label{bilingual}
\begin{aligned}
    \dot{x} &= - r_\st{X \to Z} \, x +  r_\st{Z \to X} \, z \, , \\
    \dot{y} &= - r_\st{Y \to Z} \, y +  r_\st{Z \to Y} \, z \,  , \\
    \dot{z} &= - \dot{x} - \dot{y}       \, .
\end{aligned}
\end{equation}
Interpreting such a three-state model in physical kinetics terms, the quantities $r_\st{X,Y \to Z}$ and $r_\st{Z \to X,Y}$ represent rate constants and the bilingual state Z plays the role of an ``activated state''.

With the exception of young children, language learning is typically a faster process than language attrition \cite{Sebastián-Gallés_2015,Higby_Obler_2015} and therefore, it is reasonable to assume that $r_\st{X, Y \to Z} \gg r_\st{Z \to X, Y}$ for comparable values of the speakers' fractions. 
To fully define the model, a concrete choice of the rates $r_\st{X, Y \to Z},\, r_\st{Z \to X, Y}$ is required. 
In analogy to the AS model \cite{Abrams-2003a}, Minett and Wang formulated their model focusing on the attractiveness of the language to be learned or maintained, assuming the following form for the transition rates,
\begin{subequations} 
\label{Lrates}
\begin{align}
    \label{Lrates-learning}
    &r_\st{X \to Z} = k_\st{XZ} \, y^a \, , \quad   r_\st{Y \to Z} = k_\st{YZ} \, x^a \, , 
    \\ 
    \label{Lrates-attrition}
    &r_\st{Z \to X} = k_\st{ZX} \, x^a \, , \quad  r_\st{Z \to Y} = k_\st{ZY} \, y^a \, .
    \end{align}  
\end{subequations}
The ``attractiveness'' of a language is supposed to represent the social pressure to shift towards that language, resulting from the combined effect of (a) the language prestige, i.e. the social status or the economic advantage of the language perceived by speakers, which is quantified by the  rate constants $k$~\cite{Minett2008a} and 
(b) the population size of the recruiting population, measured by the fraction of speakers of the language that will be learned or maintained,  with volatility parameter $a$ quantifying how speakers' tendency to change language depends on speaker fractions \cite{Vazquez2010a}. 

The MW model has four equilibrium points overall, of which only two (corresponding to consensus in X or Y language) are stable \cite{Chapel2010a,Vazquez2010a}. 
Consequently, similarly to the AS model, the MW model predicts that one of the languages eventually becomes extinct.
Which one of the languages disappears, depends on the specific values of the parameters and on the initial conditions \cite{Minett2008a,Heinsalu-2014b}.

\subsection{The role of bilinguals in learning and attrition} \label{sec:bilinguals}

The MW model described above implicitly makes a particular  choice concerning the role of bilinguals in language learning and attrition. 
To formulate a more general three-state model it is instructive to discuss this role here explicitly. 

\textit{Learning.}
It is natural to assume that the learning rates are determined by the attractiveness of a language, as in the AS model.
In fact, following the AS model, the MW model assumes that language attractiveness depends on the number of monolinguals only and not on that of bilinguals. 
However, one may argue that bilinguals can also contribute to monolinguals learning the other language.
A more general form of learning rates, $r_\st{X \to Z} \propto (y_{\eff}^+)^a$ and $r_\st{Y \to Z} \propto (x_{\eff}^+)^a$, in place of the learning rates \eqref{Lrates-learning}, was suggested in Ref.~\cite{Heinsalu-2014b}. 
The effective fractions of Y and X speakers  $y_{\eff}^+$ and $x_{\eff}^+$ include contributions from the bilinguals,
\begin{equation}
    \begin{aligned} \label{eq:xy_eff}
        &y_{\eff}^+ = y + \beta_Y^+ z \, ,\\
        &x_{\eff}^+ = x + \beta_X^+ z \, ,
    \end{aligned}
\end{equation}
where the specific values of the constants $\beta_X^+$ and $\beta_Y^+$ might depend on the historical and social situation.
This form is suited to describe both the effects of social pressure and those of linguistic interactions: for example, $\beta_Y^+$ can vary between $\beta_Y^+=0$, when bilinguals adapt to the language of X-monolinguals or do not exert any social pressure on the adoption of language Y, and $\beta_Y^+=1$, when bilinguals do not use language X with X-monolinguals or exert a highest social pressure in favor of language Y.
To summarize, the general case can be described through learning rates and corresponding effective fractions of speakers given by
\begin{subequations}
\label{Lrates-1}
\begin{align}
    \label{Lrates-learning-1}
    &r_\st{X \to Z} = k_\st{XZ} \, \left(\,y_{\eff}^+\,\right)^a,
    &r_\st{Y \to Z} = k_\st{YZ} \, \left(\,x_{\eff}^+\,\right)^a,
    \\
    \label{Lrates-fractions-1}
    &y_{\eff}^+ (y,z) = y + \beta_Y^+\, z,
    &x_{\eff}^+ (x,z) = x + \beta_X^+\, z .
\end{align}
\end{subequations}
%
%
\textit{Attrition.}
The attrition process, in turn, is not determined exclusively by the attractiveness of the language that will be maintained, as Eqs.~\eqref{Lrates-attrition} seem to imply; this is only one side of the coin.
The other side is that the loss of a language by attrition takes place if the size of the linguistic community is too small to provide a sufficiently high interaction rate to maintain the language or to counteract the social pressure favoring the other language. 
The conservation law $x + y + z = 1$ allows to rewrite the MW attrition rate for one of the languages (X for definitiveness, the consideration of Y is analogous) in a way that emphasizes the dependence on the community size of X-speakers: $r_\st{Z \to Y} \propto y^a = (1 - x_{\eff}^-)^a$, where effective fraction of X-speakers $x_{\eff}^-$ is $x_{\eff}^-(x,z) = x+z$. 
This particular choice of $x_{\eff}^-$ can be interpreted as ``bilinguals contribute as much as monolinguals to sustaining language X'', meaning, e.g. that bilinguals talk X with each other and/or value the language X similarly to monolinguals.
In Ref.~\cite{Castello2006a} Castell\'o et al. introduced an alternative language dynamics model with bilinguals, the AB model, characterized by attrition rates that are decreasing functions of the fraction of monolingual speakers of the language being lost;
for example, in the AB model the attrition rate of language X is $r_\st{Z \to Y} \propto (1 - x)^a$; corresponding to a situation where bilinguals play no role in preservation the language neither by talking it between each other nor by creating social pressure in favor of its preservation.
The more general approximation that describes both the X- or Y-monolinguals and a fraction $\beta_X^-$ or $\beta_Y^-$ of bilinguals supporting language X or Y, respectively, can be written as 
\begin{subequations}
\label{Lrates-2}
\begin{align}
    \label{Lrates-attrition-2}
    &r_\st{Z \to X} = k_\st{ZX} \, (1 - y_{\eff}^-)^a,
    &r_\st{Z \to Y} = k_\st{ZY} \, (1 - x_{\eff}^-)^a,
    \\
    \label{eq_newrates-attrit}
    &y_{\eff}^- (x,z)= y + \beta_Y^-\,z,
    &x_{\eff}^- (x,z) = x + \beta_X^-\, z  \, .
\end{align}
\end{subequations}
Here MW model corresponds to choosing $\beta_X^-=\beta_Y^-=1$, and AB model --- to $\beta_X^-=\beta_Y^-=0$. 

Thus, the learning  \eqref{Lrates-1} and  attrition \eqref{Lrates-2} rates describe a general family of three-state language dynamics models with bilinguals, which assign different and independent roles to the bilingual community in the learning and attrition processes.
The general dependence of the rescaled learning rate $r_\st{X,Y \to Z}/k_\st{XZ,YZ}$ versus the effective fraction $y_{\eff}^+$ or $x_{\eff}^+$, for a volatility $a > 1$, is shown in Fig.~\ref{S-shape}A.
Figure~\ref{S-shape}B shows the corresponding rescaled attrition rate $r_\st{Z \to X,Y}/k_\st{ZX,ZY}$ versus the effective fractions $y_{\eff}^-,x_{\eff}^-$.

\subsection{Threshold model} \label{sec:TMW-model}

\begin{figure}[t]
\includegraphics[width=9 cm]{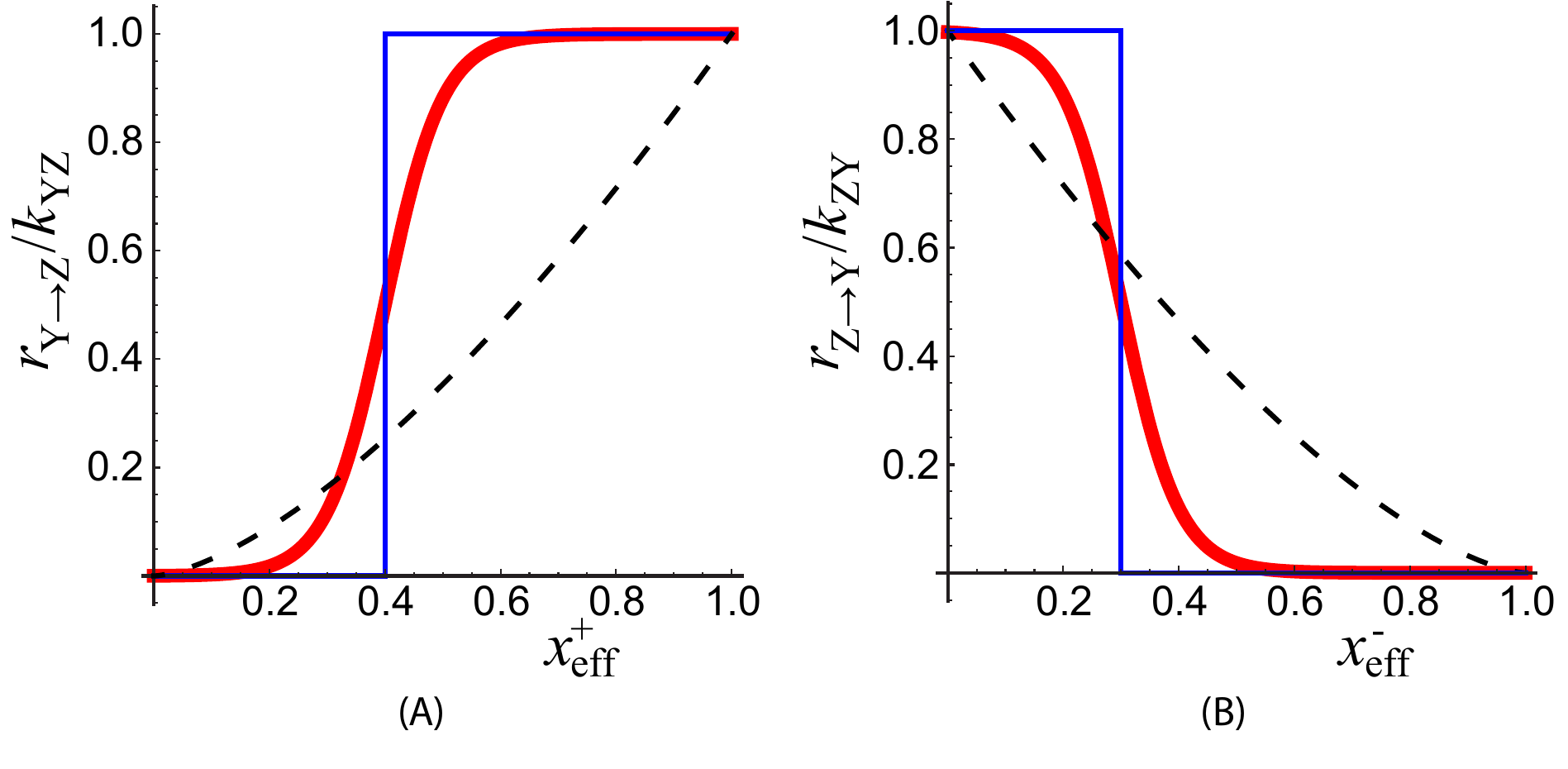}
\caption{Panel (A): normalized learning rate, $r_\st{Y \to Z}/k_{YZ}$, versus the corresponding effective fraction of X-speakers, $x_\eff^+$, needed to learn language X.
Panel (B): normalized attrition rate, $r_\st{Z\to Y}/k_{ZY}$, versus the corresponding effective fraction of X-speakers, $x_\eff^-$, needed to maintain language X. 
Different curves correspond to different three-state models:
dashed black curves---MW bilinguals model with volatility $a>1$, Eqs.~\eqref{Lrates-2};
thin blue solid curves---threshold model with step function rates, Eqs.~\eqref{rates};
thick red solid curves---heterogeneous threshold model with sigmoidal shape.
A similar scheme holds for the Y-language.
}
\label{S-shape}
\end{figure}

In the mean-field approach, memory and learning can be incorporated into the MW model, in a way similar to how it is done in the case of the AS model in Ref.~\cite{Tamm-2025}.  

The main idea of model ~\cite{Tamm-2025} is that individuals learn a new language only if they have frequent enough contacts with its speakers, making language learning a threshold process.
That is to say, if the level of exposure to a language is below the threshold, learning of this language is heavily suppressed~\cite{Scialla-2023a} either due to insufficient repetition of the learning encounters or because of low social pressure to learn the language. 
This implies that language learning rates have a sigmoid form with a \textit{lower} threshold --- rather than a power-law form --- as a function of the (effective) fraction of speakers of the learned language;
as a first approximation, the learning rate can be modeled by a step function~\cite{Tamm-2025}, see \fig{S-shape}A. 

Besides language learning, the MW model includes also language attrition. 
Following the same logic, we suggest that if the exposure of a bilingual to one of the languages is above some critical value, this language will not be forgotten, while if the exposure is below the critical value, attrition takes place and the bilingual will eventually become a monolingual, in a way similar to the standard MW model.
Thus, the attrition rates are expected to take a sigmoid form with an \textit{upper} threshold on the (effective) fraction of speakers of the language to be maintained --- which once again can be approximated as a step function, see \fig{S-shape}B.

%
%
%


Following Ref.~\cite{Tamm-2025}, introducing the thresholds in the learning rates, Eqs.~\eqref{Lrates-1}, and the attrition rates, Eqs.~\eqref{Lrates-2}, we obtain
\begin{equation} \label{rates}
\begin{aligned}
    r_\st{X\to Z} & = k_\st{XZ} \, \Theta\!\left( y_{\eff}^+ (y,z)- y_+^* \right), \\
    r_\st{Y\to Z} & = k_\st{YZ} \, \Theta \! \left(x_{\eff}^+ (x,z)- x_+^* \right), \\
    r_\st{Z\to X} & = k_\st{ZX} \, \Theta \!\left(-y_{\eff}^- (y,z)+ y_-^* \right), \\
    r_\st{Z\to Y} & = k_\st{ZY} \, \Theta \! \left(-x_{\eff}^- (x,z)+ x_-^* \right)\, ,
\end{aligned}
\end{equation}
where $ \Theta (x)$ is the Heaviside $\Theta$-function, the effective fractions of $x,y$ speakers $x_{\eff}^{\pm}, y_{\eff}^{\pm}$ are given by  \eqref{Lrates-fractions-1},\eqref{eq_newrates-attrit} and the parameters $x_{\pm}^*, \, y_{\pm}^*$ represent their threshold values. 
Each of the conditions at which Heaviside functions in \eqref{rates} switch from zero to one,
\begin{equation} \label{conditions}
\begin{aligned}
    y_{\eff}^+ (y,z)= y_+^* , \quad
    x_{\eff}^+ (x,z)= x_+^* , \\
    y_{\eff}^- (y,z)= y_-^* , \quad
    x_{\eff}^- (x,z)= x_-^* \, ,
\end{aligned}
\end{equation}
represents a straight line in the ternary diagram, splitting it into two regions with different values of the corresponding Heaviside function, see \fig{ternary}C. 



 
%
%
Following Ref.~\cite{Scialla-2023a}, we expect bilinguals to be more relevant for language retention than for language learning,
\begin{equation} \label{eq:beta}
    \beta_X^- \geq \beta_X^+,   \quad     \beta_Y^- \geq \beta_Y^+.
\end{equation}
Indeed, monolingual-bilingual encounters occur mainly in the language both speakers know, providing relatively few opportunities for monolinguals to learn the second language, while for bilinguals, encounters with other bilinguals are an important incentive to sustain an otherwise endangered language~\cite{Sanchez-2021}. 
As a result, separation lines  on the ternary diagram corresponding to retention are steeper than those corresponding to learning, see \fig{ternary}D.
If the assumption \eqref{eq:beta} holds, we do not expect the results to be very sensitive to a particular choice of $\beta_{X,Y}^\pm$. 
In what follows, we assume for definiteness
\begin{equation}
    \beta_{X,Y}^- =1/2, \quad   \beta_{X,Y}^+ = 0,
\end{equation}
which implies that for learning purposes bilinguals are irrelevant, i.e., presumably all communications happen in mutually understandable language and social pressure from bilinguals does not play a role, while for the purpose of retention of a given language each bilingual provides half of the corresponding monolingual contribution. 
That is to say, using these conditions in Eqs.~\eqref{rates} and \eqref{Lrates-2}, we will consider the system
\begin{equation} 
    \begin{aligned}
        \dot x &= -\kappa \, x \, \Theta (y - y_+^*) + \gamma_x \, z \, \Theta (y_-^*-y - z/2), \\
        \dot y &= - (1-\kappa) \, y \, \Theta (x - x_+^*) + \gamma_y \, z \, \Theta (x_-^*-x - z/2), \\
        \dot z &= - \dot x-\dot y ,
    \end{aligned}
    \label{system}
\end{equation}
%
In these equations, we renormalized time, $t \rightarrow (k_\st{XZ} + k_\st{YZ}) t$  and introduced dimensionless parameters $\kappa = k_\st{XZ}/(k_\st{XZ} + k_\st{YZ})$ and $\gamma_\st{x, y} = k_\st{ZX, ZY}/(k_\st{XZ} + k_\st{YZ})$. This means that the unit of time is around 10 years, while $\gamma_\st{x, y}$, being the ratios of attrition to learning rates, are assumed to be small, typically of order $10^{-1}$.

In what follows, by ``symmetric case'' we refer to the parameter set that defines the two languages as being completely equivalent to each other in terms of learning difficulties and maintenance efforts, being characterized by the same transition rates ($\kappa=1-\kappa=1/2, \gamma_x = \gamma_y \equiv \gamma$), as well as by equal learning and attrition thresholds, ($x_+^* = y_+^* \equiv s_+\, , \quad x_-^* = y_-^* \equiv s_-$), so that the set of equations reduces to
\begin{equation}
\begin{aligned}
    \dot x &= -x \, \Theta (y - s_+) + \gamma \, z \, \Theta (s_--y - z/2), \\
    \dot y &= -y \, \Theta (x - s_+) + \gamma \, z \, \Theta (s_--x - z/2), \\
    \dot z &= - \dot x - \dot y .
\end{aligned}
\label{system_sym}
\end{equation}
%

%
\begin{figure*} 
\includegraphics[width=17cm]{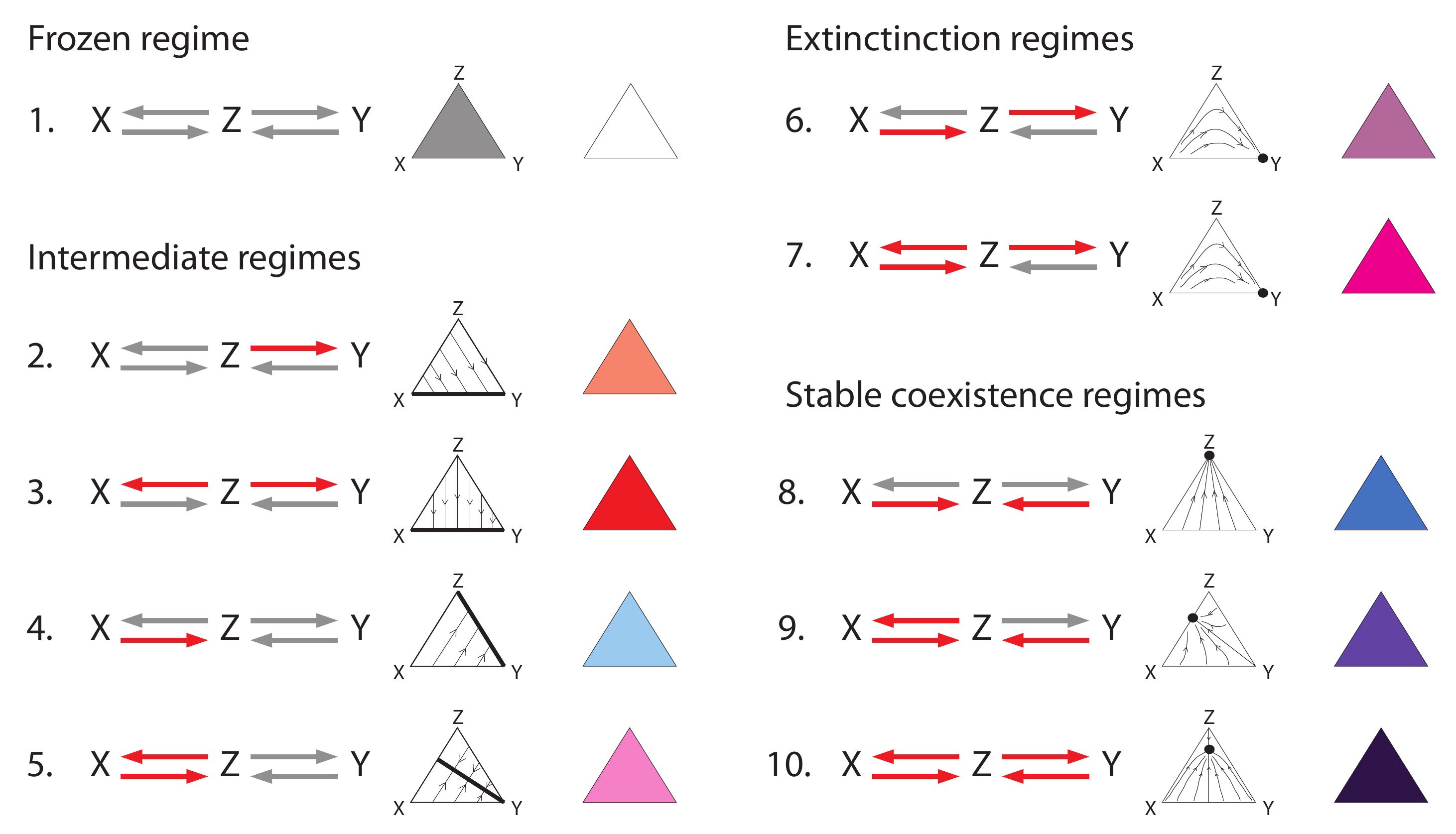}
\caption{Dynamics of system \eqref{system} in the case when each reaction channel is either permanently open (red arrow) or permanently closed (gray arrow). 
The bold dots and lines on the ternary plots indicate the attractors. Thin lines correspond to a schematic representation of evolution trajectories and arrows on them indicate the direction of the evolution. 
Gray area is the area of neutral stability. 
The exemplary trajectories are drawn for a fully symmetric case $\gamma_x = \gamma_y$, $\kappa = 1/2$. 
The last column introduces color-coding of the states, which is used throughout the following figures. 
All states except 1, 3, 7 and 10 have dual counterparts, which can be obtained by $X \longleftrightarrow Y$ replacement and are not shown to save the space.}
\label{fig:pure}
\end{figure*}
%

\section{Results} \label{sec:results}

\subsection{Pure regimes}
\label{sec:pure}

At any time $t$, each Heaviside function in Eqs.~\eq{system}, depending on the values of the population fractions $x(t), y(t), z(t)$  can take either value 0 or 1, meaning that the correspondent  process is switched off or on, respectively.
Altogether, for each combination of $x,y,z$ there are $2^4=16$ combinations of possible values of the Heaviside functions and therefore 16 different possible dynamical regimes. 
Generally, a trajectory of the system is associated with more than one of these dynamical regimes: 
while $x(t)$ and $y(t)$ change, the values of corresponding Heaviside functions might switch between 0 and 1. 

However, it is instructive to start the study of Eqs.~\eq{system} by considering the situations in which each of the four Heaviside functions present in Eqs.~\eq{system} is resolved in the same way for all values of $x,y,z$, i.e., where all four functions are always 0, where the first one is always 1 and the other three are always 0, etc. 
We refer to such situations as ``pure regimes'' of Eqs.~\eq{system}, as opposed to the more general mixed regimes, where Heaviside functions are resolved differently for different values of $x,y,z$. 
Indeed, the pure dynamic regimes essentially are the building blocks from which the solutions of mixed regimes can be constructed.  

Clearly, all the pure regimes of  Eqs.~\eq{system} are linear and thus easy to solve. 
We summarize the results for each  regime in the list below and visualize them in \fig{fig:pure}, which contains the list of on/off processes for each regime and corresponding ternary flow diagram. 
For brevity, we omit the regimes, which can be obtained from regimes 2, 4, 5, 6, 7 and 9 by replacing $x \leftrightarrow y, \kappa \leftrightarrow (1 \!-\! \kappa)$. 
We start with the regimes 1, 6-10, which are analogous to those observed in Ref.~\cite{Tamm-2025}, and finish with the new class of intermediate regimes 2-5.

\subsubsection{Frozen regime}

{\it Regime 1.} 
This regime corresponds to a completely frozen state
\begin{equation}
 \dot{x} = \dot{y} = 0 \, , 
\end{equation} 
in which nothing changes with respect to the initial configuration:
\begin{equation}
    x(t) = x_0\, , \quad y(t) = y_0\, , \quad z(t) = z_0 \, .
\end{equation} 

\subsubsection {Extinction regimes}

There are two regimes, 6 and 7, that correspond to eventual extinction of one of the languages.

{\it Regime 6.} 
The dynamic equations in this regime read
\begin{equation}
\begin{aligned}
   \dot{x} &= -\kappa x \, , \\ 
   \dot{z} &= \kappa x- \gamma_y z \, .
   \end{aligned}
\end{equation} 
The solution is a linear combination of two decaying exponents:
\begin{equation}
\begin{aligned}
    x(t) &= x_0 \exp(-\kappa t) \, , \\ 
    y(t) &= 1 - x(t)-z(t) \, , \\
    z(t) &= \left (z_0 + \frac{\kappa x_0}{\kappa-\gamma_y} \right) \, \exp (-\gamma_y t) -  \frac{\kappa x_0}{\kappa-\gamma_y} \exp(-\kappa t) \, ,  
  \end{aligned} 
\end{equation} 
except for the (physically unrealistic) case $\kappa = \gamma_y$, when the solutions take the form $(a+b t) \exp (-\kappa t)$ . 
In this case all monolinguals of language X and all bilinguals eventually turn into monolinguals of language Y. The process includes fast learning mode with relaxation time $1/\kappa$ and slow attrition mode with relaxation time $1/\gamma_y$. 
As a result, in the beginning of the time evolution, the number of bilinguals might grow due to the X $\to$ Z process, but eventually decays at larger times, in a way similar to the solutions of the MW model.

{\it Regime 7.} 
This regime is similar to the previous one, with the following dynamic equations:
\begin{equation}
\begin{aligned}
   \dot{x} &= -\kappa x + \gamma_x z \, , \\ 
   \dot{z} &= \kappa x - (\gamma_x + \gamma_y)z \, .
  \end{aligned} 
\end{equation} 
Structurally, the solutions are once again a combination of two exponents: 
\begin{equation}
\begin{aligned}
   & x(t) = A_1 \exp(-t/\tau_1) + A_2 \exp(-t/\tau_2) \, ,   \\
   & z(t) = B_1 \exp(-t/\tau_1) + B_2 \exp(-t/\tau_2) \, , 
   \end{aligned}
\end{equation} 
where
\begin{equation*}
   \tau_{1,2} = \displaystyle \frac{\gamma_x + \gamma_y + \kappa \pm \sqrt{(\gamma_x + \gamma_y + \kappa)^2 - 4 \gamma_y \kappa}}{2 \gamma_y \kappa} \, .
\end{equation*} 
The exact expressions for $A_{1, 2}$ and $B_{1, 2}$ in terms of initial conditions is straightforward to compute but are rather cumbersome. 

Note that in the limit $\kappa \gg \gamma_{x,y}$ it follows that $\tau_2  \gg \tau_1$. 
This means that in this limit, similarly to regime 6, the dynamics has two distinct stages. Initially, at times of order $\tau_1 \sim 1$, monolinguals Y disappear and a partial equilibrium between X-monolinguals and bilinguals is reached, and then this dynamic monolingual-bilingual mix slowly forgets language Y and turns into monolinguals with relaxation time $\tau_2 \sim  \gamma$. This dynamics once again leads to intermittent growth in the number of bilinguals for $\tau_1<t<\tau_2$. 
%

\subsubsection{Stable coexistence regimes}

The regimes 8, 9, and 10 correspond to long-term language coexistence with a single attractive stable point. The outcome in theses regimes is independent of the initial conditions. 

{\it Regime 8.} 
In this regime monolinguals of both languages learn the other language and become bilinguals,
\begin{equation}
\begin{aligned}
    \dot{x} &= -\kappa x \, , \\ 
    \dot{y} &= (1-\kappa)y \, ,
    \end{aligned}
\end{equation}
and the final state is a bilingual society:  
\begin{equation}
\begin{aligned}
    x(t) &= x_0 \exp(- \kappa t) \, , \\ 
    y(t) &= y_0 \exp[-(1 - \kappa) t] \, , \\
    z(t) &= 1 - x(t) - y(t).    
  \end{aligned} 
\end{equation}

{\it Regime 9.} 
This regime is described by the following dynamic equations:
\begin{equation}
\begin{aligned}
    \dot{x} = -\kappa x + \gamma_x z \, , \\ 
   \dot{y} = -(1 - \kappa) y \, .
   \end{aligned}
\end{equation}
In this regime, also in the long time limit, there is a finite fraction of monolinguals X present in the system:
\begin{equation}
\begin{aligned}
    x(t) &= x_\infty + A_1 \exp(-t/\tau_1) + A_2 \exp(-t/\tau_2) \, , \\
    y(t) &= y_0 \exp(-t/\tau_2) \, ,  \\ 
    z(t) &= 1 - x(t) - y(t) \, , \\
  \end{aligned} 
\end{equation}
with
\begin{equation*}
     x_\infty = \displaystyle \frac{\gamma_x}{\gamma_x + \kappa} \, , \quad \tau_1 =  \frac{1}{\gamma_x + \kappa} \, , \quad \tau_2 =  \frac{1}{1 - \kappa} \, .  
\end{equation*}
It is straightforward to express  $A_{1, 2}$ in terms of initial conditions. 
Furthermore, although the solution once again is a combination of two exponents now, unlike in the extinction regimes 6 and 7, there is no separation of timescales: both relaxation times are of order 1. 

{\it Regime 10.} 
This is the case where all processes are allowed and the dynamic equations read:
\begin{equation}
\begin{aligned}
    \dot{x} &= -\kappa x + \gamma_x z \, , \\ 
    \dot{y} &= -(1 - \kappa)y + \gamma_y z \, . 
    \end{aligned}
\end{equation}
In equilibrium, the fractions of all three linguistic groups remain fraction: 
%
\begin{equation}
\begin{aligned}
    & x(t) = x_\infty + A_1 \exp(-t/\tau_1) + A_2 \exp(-t/\tau_2) \, , \\
    & y(t) = y_\infty + B_1 \exp(-t/\tau_1) + B_2 \exp(-t/\tau_2) \, , \\
    & z(t) = 1 - x(t) - y(t) \, , 
    \end{aligned}  \label{regime10}
\end{equation}
with
\begin{equation*}
\begin{aligned}
    & x_\infty = \displaystyle \frac{\gamma_x (1 - \kappa)}{\gamma_x (1 - \kappa) + \gamma_z \kappa + \kappa (1 - \kappa)} \, , \\
    & y_\infty = \displaystyle \frac{\gamma_y \kappa}{\gamma_x (1 - \kappa) + \gamma_z \kappa + \kappa (1 - \kappa)} \, , \\
    & \tau_{1, 2} =  \displaystyle \frac{2}{1 \pm \sqrt{(1 - 2\kappa)^2 + \gamma_x \gamma_y}} \, .  
  \end{aligned}
\end{equation*}
Both relaxation times $\tau_{1, 2}$ are of order 1, except for the vicinity of the point
\begin{equation}
    \kappa_\st{res} = \left(1 \pm \sqrt{1 - \gamma_x\gamma_y}\right)/2 \, ,
\end{equation} 
where $\tau_2$ diverges. 
However, for slow language attrition, $\gamma_x \gamma_y \ll 1$ this condition implies unrealistically large asymmetry in learning rates. 
The expression of numerical parameters $A_{1, 2}, B_{1, 2}$ in terms of initial condition is straightforward but rather cumbersome.

\subsubsection{Intermediate regimes}

Regimes 2, 3, 4, and 5 provide the most interesting dynamics. 
In these regimes, the attractor is not a single point but rather a line of points on the ternary diagram, so that the exact outcome, i.e., the point to which the dynamics converges, depends on the initial conditions. In terms of stability analysis, each of the points of the attractor can be called ``partially neutral''. That is to say, they are attractive with respect to infinitesimal perturbation in one direction, along the trajectories in \fig{fig:pure}, and neutral (neither attractive nor repulsive) with respect to infinitesimal perturbation in the direction perpendicular to the trajectories.
In other words, in each of these regimes there is an integral of motion which is conserved throughout the dynamics, so that different points on the attractor correspond to different values of this integral of motion.

{\it Regime 2.} 
In this regime the fraction of monolingual speakers of language X is conserved:
\begin{equation}
\begin{aligned}
    &\dot{x} = 0 \, , \\ 
    &\dot{y}=\gamma_y z \, .
    \end{aligned}
\end{equation} 
Solving these equations, we see that 
\begin{equation}
\begin{aligned}
    & x(t) = x_0 \, , \\ 
    &y(t) = y_0 + z_0 \left[1- \exp (-\gamma_y t) \right] \, , \\
    & z(t) = z_0 \exp(-\gamma_y t)  \, ,
\end{aligned}
\end{equation} 
i.e., all bilinguals initially present in the system eventually become monolinguals of language Y.
Thus, any point on the XY side of the ternary diagram can be the final state of the dynamics, depending on the initial fraction $x_0$ of the monolingual speakers of language X.

{\it Regime 3.} 
Dynamics in this regime is described by
\begin{equation}
\begin{aligned}
  &\dot{x}=\gamma_x z \, , \\ 
  &\dot{y}=\gamma_y z \, , 
  \end{aligned} \label{regime3eq}
\end{equation} 
so that 
  $\gamma_y x - \gamma_x y = \mathrm{const}$.
is the conserved quantity. 
The solutions of the dynamic equations (\ref{regime3eq}) read,
\begin{equation}
\begin{aligned}
 & x(t) = x_0 + \frac{\gamma_x}{\gamma_x + \gamma_y} z_0 \left\{1 - \exp [-(\gamma_x + \gamma_y) t]\right\} \, ,  \\ 
 & y(t) = y_0 + \frac{\gamma_y}{\gamma_x + \gamma_y} z_0 \left\{1 - \exp [-(\gamma_x + \gamma_y) t]\right\} \, , \\
  & z(t) = z_0 \exp [-(\gamma_x + \gamma_y) t ] \, .
\end{aligned}
\end{equation} 
Thus, similarly to the previous regime, the bilinguals become extinct and the final outcome is a point on the XY side of the ternary diagram. 
Which particular point is chosen, depends on the value of $\gamma_y x_0 - \gamma_x y_0$, i.e., also on the initial fractions of X and Y speakers.

{\it Regime 4.} 
In this regime 
\begin{equation}
\begin{aligned}
     &\dot{x} =-\kappa x \, , \\ 
     &\dot{y}=0 \, , 
     \end{aligned} \label{regime4eq}
\end{equation} 
i.e., the fraction of monolinguals of language Y is conserved, while monolinguals of X turn into bilinguals.
The solutions of the dynamic equations (\ref{regime4eq}) read,
\begin{equation}
\begin{aligned}
    & x(t) = x_0 \exp(-\kappa t) \, , \\ 
    &y(t) = y_0 \, , \\
    & z(t) = z_0 + x_0 \left[1- \exp(-\kappa t)\right] \, ,
\end{aligned} 
\end{equation} 
i.e., the outcome is a point on the YZ side of the ternary diagram determined by the initial value $y_0$.

{\it Regime 5.} 
Here, similarly to the previous regime, the fraction of monolinguals of language Y is conserved, 
%
\begin{equation}
\begin{aligned}
    &\dot{x} = -\kappa x + \gamma_x z, \\ 
    &\dot{y} = 0 \, .
    \end{aligned} \label{regime5eq}
\end{equation}
The solution of the dynamic equations (\ref{regime5eq}) reads,
\begin{equation}
\begin{aligned}
    & x(t) = x_\infty + (x_0 - x_\infty) \exp(- t/\tau) \, , \\
    & y(t) = y_0 \, , \\ 
    &z(t) = 1 - x(t) - y(t) \, , 
    \end{aligned} 
\label{regime5}
\end{equation}
with
\begin{equation*}
    x_\infty = \displaystyle (1 - y_0)\frac{\gamma_x}{\gamma_x + \kappa} \, , \quad \tau =  \frac{1}{\gamma_x + \kappa} \, . 
\end{equation*}
The attractor in this case is a line connecting the Y corner of the diagram and a point splitting the XZ side in proportion $\gamma_x : \kappa$, with the exact point determined by the initial fraction $y_0$ of monolingual speakers of language Y.
%


%
\begin{figure*}
\centering
\includegraphics[width=15cm]{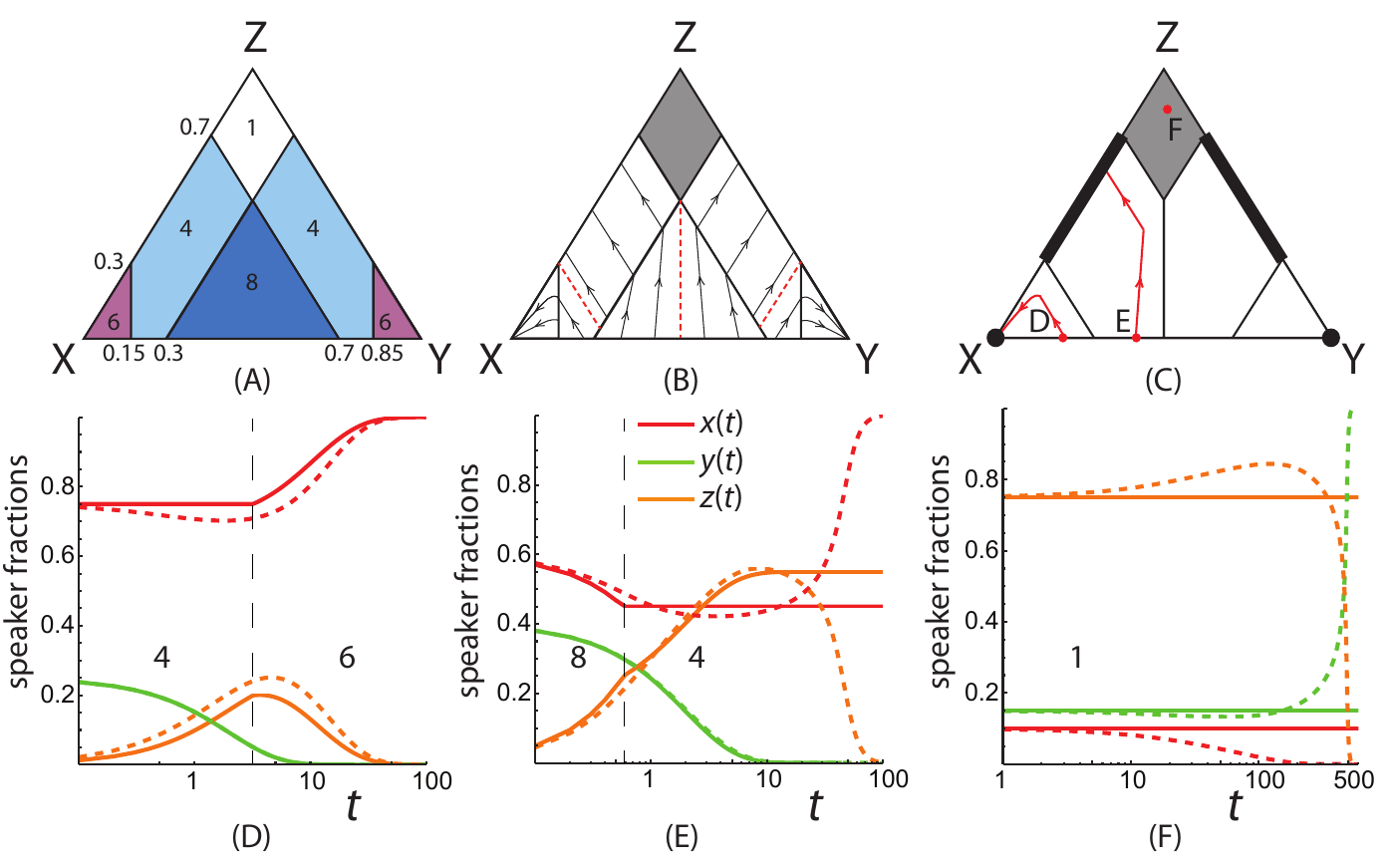}
\caption{Example of dynamics with $x^*_+ =y^*_+= 0.3, x^*_- =y^*_- = 0.15, \kappa = 1/2$. 
(A) The switching conditions split the whole space of possible compositions into 7 areas -- numbers and colors correspond to the pure states introduced in \fig{fig:pure} and in the text; 
(B) schematic depiction of the dynamic flows in the system -- red dashed lines separate basins of attraction; 
(C) attractors (black bold lines and points) of Eqs.~\eq{system} for this particular set of parameters and boundaries of their basins of attraction; red dots and lines correspond to the initial conditions and trajectories of the dynamics shown in panels (D)-(F); 
(D)-(F) examples of dynamics for three different sets of initial conditions for $\gamma_{x,y}=0.1$; red, green and orange lines show the time evolutions of $x(t),y(t),z(t)$ given by \eqref{system} with Heaviside (full lines) and S-shaped (\eqref{sshape}, dashed lines) rates; the dashed vertical lines indicate the switching between the regimes, with the numbers inside the panels indicating the regimes;
(D) initial condition $(x_0,y_0,z_0)=(0.75,0.25,0)$ converging to the extinction of language Y, $(x_\infty,y_\infty,z_\infty)=(1,0,0)$;
(E) evolution starting from initial condition $(x_0,y_0,z_0)=(0.6,0.4,0)$ converging to a long-term coexistence of monolinguals of X and bilinguals , $(x_\infty,y_\infty,z_\infty)=(0.45,0.55,0)$;, 
(F) evolution starting from initial condition $(x_0,y_0,z_0)=(0.1,0.15,0.75)$ is frozen from the start, so $(x_\infty,y_\infty,z_\infty)=(0.1,0.15,0.75)$.}
\label{example}
\end{figure*}
%

\subsection {An example of the dynamics} \label{sec:example}

Knowledge of the system dynamics in the pure states, as described above, is sufficient to obtain a solution of Eqs.~\eqref{system} for any given set of parameters. Indeed, for any fixed set of parameters, the dynamics of Eqs.~\eqref{system} is nothing but a combination of solutions 1-10, glued together at the points where Heaviside functions pass their respective thresholds.

\begin{figure*}
\centering
\includegraphics[width=14cm]{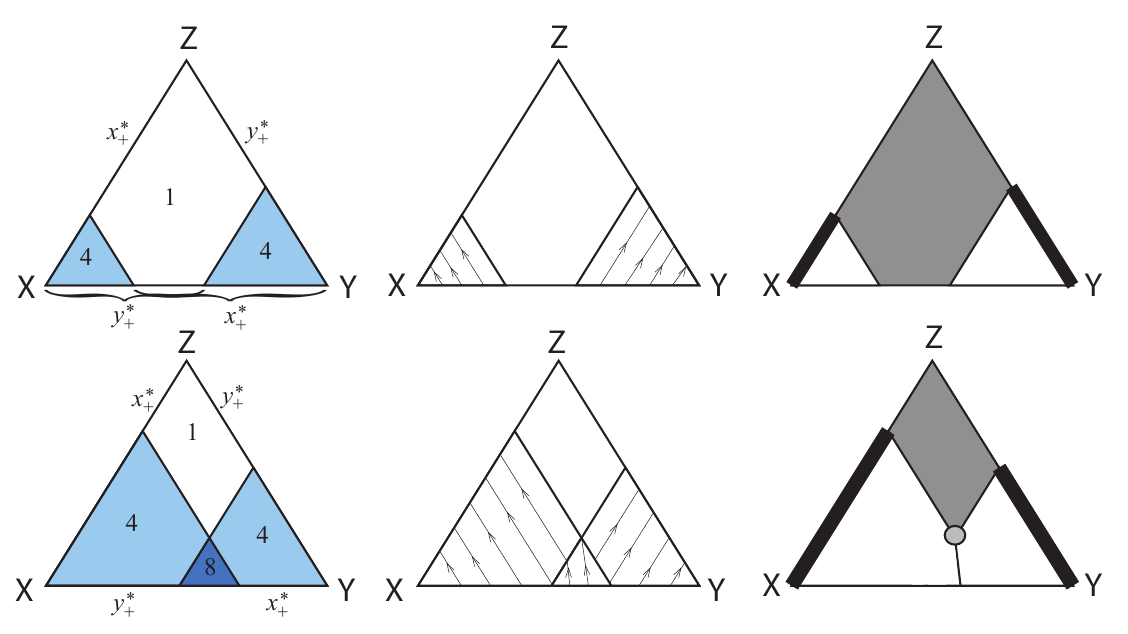}
\caption{Flows and attractors of the system with no language attrition. Top row: $x^*_+ + y^*_+>1$; bottom row $x^*_+ + y^*_+ \leq1$. First column shows splitting of the space of parameters into regions where the conditions in Eqs.~\eqref{system_noattr} are opened differently; second column is a flaw diagram; third column shows the attractors of the dynamics. Note that the shape of the line splitting the basins of attraction in the bottom raw is, generally speaking, $\kappa$-dependent and it is a straight line going through $z=1$ angle of the ternary diagram only if $\kappa=1/2$. The gray point inside the triangle in the bottom row corresponds to a partially stable stationary point at $(x,y,z)=(x_+^*,y_+^*,1-x^*_+-y_+^*)$, it is attractive along the vertical black line, repulsive in perpendicular direction, neutral if approached from the gray area side.}
\label{fig:noattr}
\end{figure*}

As an example, consider a fully symmetric system with $\kappa = 1/2$, $x^*_+ = y^*_+ = 0.3$, $x^*_- = y^*_- = 0.15$.  This means that if more than 30\% of the population are monolinguals of one or the other language, this language will be learned by the monolinguals of the other language, but if the effective fraction of the speakers of one of the languages is below 15\%, this language will be forgotten by bilinguals [see Eqs.~\eqref{eq:xy_eff}, \eqref{rates}, \eqref{eq:beta}].
The conditions at which the learning and forgetting processes switch on,
\begin{equation} \label{thresholds}
\begin{aligned}
    x &= x^*_+ = 0.3 \, , \\
    y &= y^*_+ = 0.3 \, ,  \\
    x + z/2 &= x^*_- =0.15 \, , \\
    y + z/2 &= y^*_- =0.15  \, ,
\end{aligned}
\end{equation}
split the ternary diagram into 6 sub-regions as shown in \fig{example}A. 
In each sub-region the dynamics is described by one of the pure states as color-coded in \fig{example}A, 
i.e., at the top, in the vicinity of $z = 1$ (white region, regime 1 of the previous section), all processes switch off and the dynamics is frozen; 
in the lower middle part (blue region, regime 8), both learning processes are on, thus, the monolingual groups of both languages become bilinguals; 
in the vicinity of $y = 1$ (purple region, regime 6), X monolinguals learn language Y, while bilinguals become monolinguals of language Y, etc. 
The thin lines with arrows in \fig{example}B schematically show the direction of time evolution within each sub-region; at the boundaries of the regions the evolution lines join continuously. 
Depending on the initial conditions, the evolution of the system can lead to one of the three possible outcomes (see \fig{example}C and the corresponding exemplary time dependencies $x(t)$, $y(t)$, $z(t)$ for the three regimes shown in full lines in Figs.~\ref{example}D-\ref{example}F): 
(i) evolution is frozen at the initial conditions (Fig.\ref{example}F);
(ii) complete extinction of one of the two languages (Fig.\ref{example}D); 
(iii) monolinguals of one of the two languages get extinct, and a neutral equilibrium between the monolinguals of the other language and bilinguals is reached (Fig.\ref{example}E); the size of the remaining monolingual population depends on the initial conditions and can be anything from $2x^*_- = 0.3$ to $1-x^*_+ = 0.7$.
The boundaries of the basins of attraction are shown by red dashed lines in \fig{example}B and by thin black lines in \fig{example}C.  
Note that the attrition rates $\gamma_\st{x, y}$ are completely irrelevant for the eventual outcome of the dynamics, as the only feature related to $\gamma$ is the relaxation time towards language extinction in regime (ii). 

In order to understand how the results will change in the case when there is some heterogeneity in the position of learning and attrition thresholds for different speakers, we replaced the step functions $\Theta(x-x^*)$ in Eqs.~\eqref{system} with S-shaped functions $S(x,x^*,w)$ of the form
\begin{equation}
\begin{aligned}
    S(x,x^*,w) &= \frac{T\left((x-x^*)/w\right)-T\left(-x^*/w\right)}{T\left((1-x^*)/w\right)-T\left(-x^*/w\right)}  \\
    T(x) &= \frac{1}{\pi} \arctan x+\frac{1}{2}
\end{aligned} 
\label{sshape}
\end{equation}
where $w$ has the meaning of width of the curve, $S(x,x^*,w) \to \Theta(x-x^*)$ when $w\to 0$.

In \fig{example} D-F with dashed lines we show the numerical solutions of ~\eqref{system} with such S-shaped rates for $w=0.1$. In the extinction regime (\fig{example}D) replacing the $\Theta$-functional rates with S-shaped ones leads only to minor quantitative changes in the dynamic. In the cases when $\Theta$-functional rates lead to fully (\fig{example}F) or partly (\fig{example}E) frozen dynamics the changes are more profound. In both cases, the dynamics with S-shaped rates leads to the eventual extinction of the minority language, but the time it takes is of order hundreds, i.e., well above the observational timescales (recall that the unite of time is around 10 years). Meanwhile, stable states of the system with $\Theta$-functional thresholds survive as long-living intermediate states in the case of S-shaped ones. 

Indeed, in \fig{example}E the coexistence between monolinguals of the majority language and bilinguals, which survives {\it ad infinitum} for $\Theta$-functional rates, exists as an intermediate states for $5 \lesssim t \lesssim 50$, i.e., for several centuries. In \fig{example}F the evolution is essentially frozen up to $t\approx 30$, then monolinguals of the minority language (X in this case) disappear and for $30 \lesssim t \lesssim 300$ there is a long-living intermediate state consisting of bilinguals and monolinguals Y.

We conclude that although neutral and partially neutral equilibria existing in the solutions of Eqs.~\eqref{system} do not, strictly speaking, survive the smearing out of the sharp thresholds, they are still observable as long-living transient states, and the timescale of their decay can easily be beyond the accessible observation times.

\subsection{Limiting cases}\label{sec:limiting}

%
\begin{figure*}
\centering
\includegraphics[width=14cm]{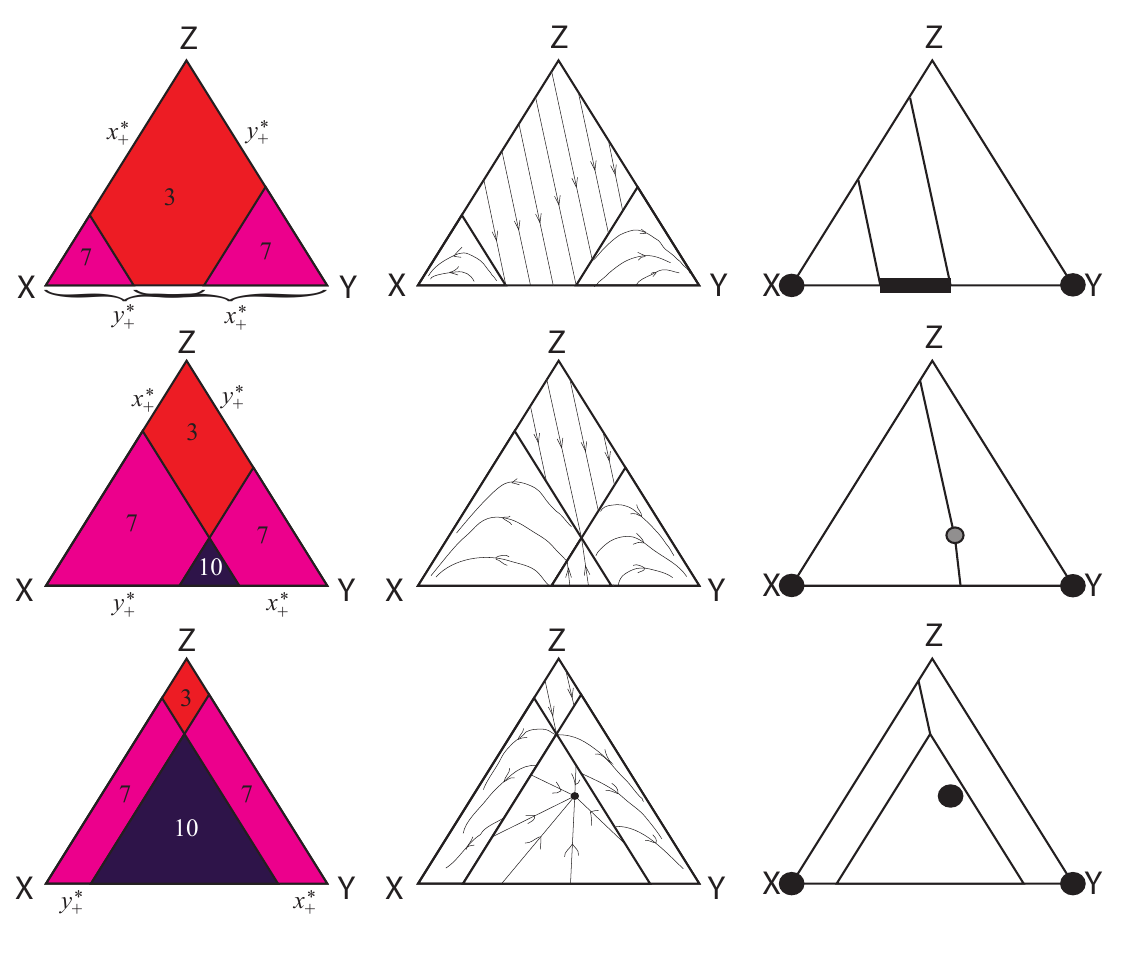}
\caption{Flows and attractors of the system where language attrition is always on. The three rows correspond to three distinct cases with respect to the values of $x^*_+, y^*_+$. 
First row: $x^*_+ + y^*_+ > 1$; second and third rows: $x^*_+ + y^*_+ \leq 1$. 
The third row corresponds to the case when the stable point is inside the dark purple triangle, i.e. $x^*_+>x_{\infty},y^*_+>y_{\infty}$, see Eqs.~\eqref{regime10cond}, the second row -- to the case when it is not, i.e., either  $x^*_+<x_{\infty}$ or $y^*_+<y_{\infty}$. 
Similarly to \fig{fig:noattr} the columns correspond to  (i) splitting of the space of parameters into regions, (ii) flaw diagram,  and (iii) attractors and basins of attraction. 
The slope of trajectories in the red region 3 is determined by the ratio $\gamma_x/\gamma_y$ (vertical if it equals 1). 
In a fully symmetric region the line separating basins of attraction in the second row is vertical. 
The gray dot in this diagram designates a saddle point at $(x,y,z)=(x_+^*,y_+^*,1-x^*_+-y_+^*)$. 
}
\label{fig:attr}
\end{figure*}
\begin{figure*}
\centering
\includegraphics[width=14cm]{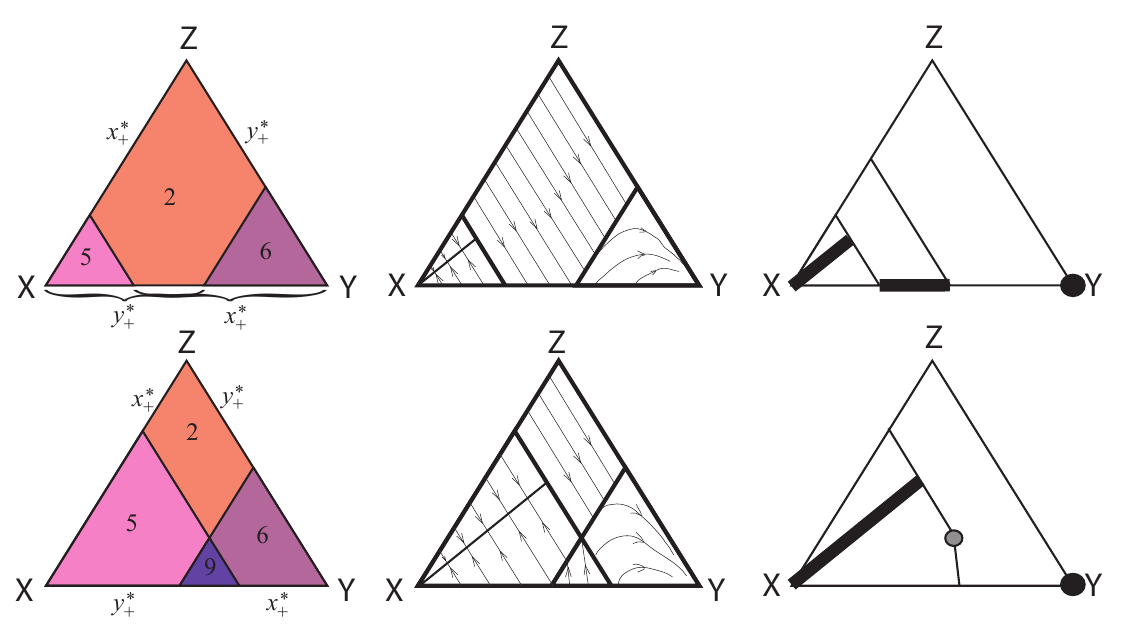}
\caption{Flows and attractors of the system where language attrition happens only for language X. First row, $x^*_+ + y^*_+>1$; second row, $x^*_+ + y^*_+ \leq1$. Similarly to the previous two figures, the columns correspond to  (i) splitting of the space of parameters into regions, (ii) flaw diagram,  and (iii) attractors and basins of attraction. The gray dot designates a partially attractive stable point at $(x,y,z)=(x_+^*,y_+^*,1-x^*_+-y_+^*)$. 
}
\label{fig:asymattr}
\end{figure*}

Notably, the set of attractors in the example described in the previous section is not only independent of the attrition rates, but  is also not very sensitive to the concrete values of $x^*_{\pm}$: a small change in any of them will result only in small shifts in the boundaries of the basins of attraction without changing the set of possible outcomes. 
Thus, it seems interesting to enumerate all the types of possible dynamics for different values of parameters in Eqs.~\eqref{system}. 
Unfortunately, a full solution of this problem seems unfeasible given that Eqs.~\eqref{system} have 7 parameters: $\kappa, \gamma_\st{x,y}, x^*_{\pm}, y^*_{\pm}$. 
However, the problem becomes traceable in several limiting cases. 
In this subsection we consider the cases, which are degenerate with respect to attrition, i.e., when attrition happens either always or never. 
In the next subsection we address the case of full symmetry between the two competing languages.

Let us start with a simple case and assume that there is no attrition, i.e., $x_-^* = y_-^* = 0$. 
Then, Eqs.~\eqref{system} become,
\begin{equation} \label{system_noattr}
\begin{aligned}
    \dot x &= -\kappa \, x \, \Theta (y - y_+^*) \, ,  \\
    \dot y &= - (1-\kappa) \, y \, \Theta (x - x_+^*) \, , \\
    \dot z &= - \dot x-\dot y \, .
\end{aligned}
\end{equation}
As shown in \fig{fig:noattr}, Eqs.~\eqref{system_noattr} always have two types of attractors: either the dynamics is completely frozen or there is what in section \ref{sec:pure} we have called an intermediate regime, resulting in a partially neutral equilibrium between one of monolingual and the bilingual populations. 
Furthermore, in the case of the frozen state there are two slightly different situations: for $x^*_+ + y^*_+\leq1$ a frozen state with no bilinguals is possible, while for $x^*_+ + y^*_+ > 1$ one needs also even a small fraction of bilinguals to freeze the dynamics.

In the case when attrition is always present, i.e., $x_-^* = y_-^* = 1$, Eqs.~\eqref{system} can be rewritten as
\begin{equation} \label{system_attr}
\begin{aligned}
    \dot x &= -\kappa x \, \Theta (y - y_+^*) + \gamma_x z \, , \\
    \dot y &= - (1-\kappa) y \, \Theta (x - x_+^*) + \gamma_y z \, , \\
    \dot z &= - \dot x-\dot y \, .
\end{aligned}
\end{equation}
The possible regimes of Eqs.~\eqref{system_attr} are shown in \fig{fig:attr} and now there are three distinct possibilities. 

First, if $x^*_+ + y^*_+ \leq 1$, there are three stable states: a partially neutral equilibrium between the two monolingual communities or consensus on one of the languages. 
If $x_0 + z_0 \gamma_x / (\gamma_x+\gamma_y) > x_+^*$ then $x = 1$, whereas if $x_0 + z_0 \gamma_x / (\gamma_x+\gamma_y) < 1-y_+^*$ then $y = 1$.

\begin{figure}
\centering
\includegraphics[width=7cm]{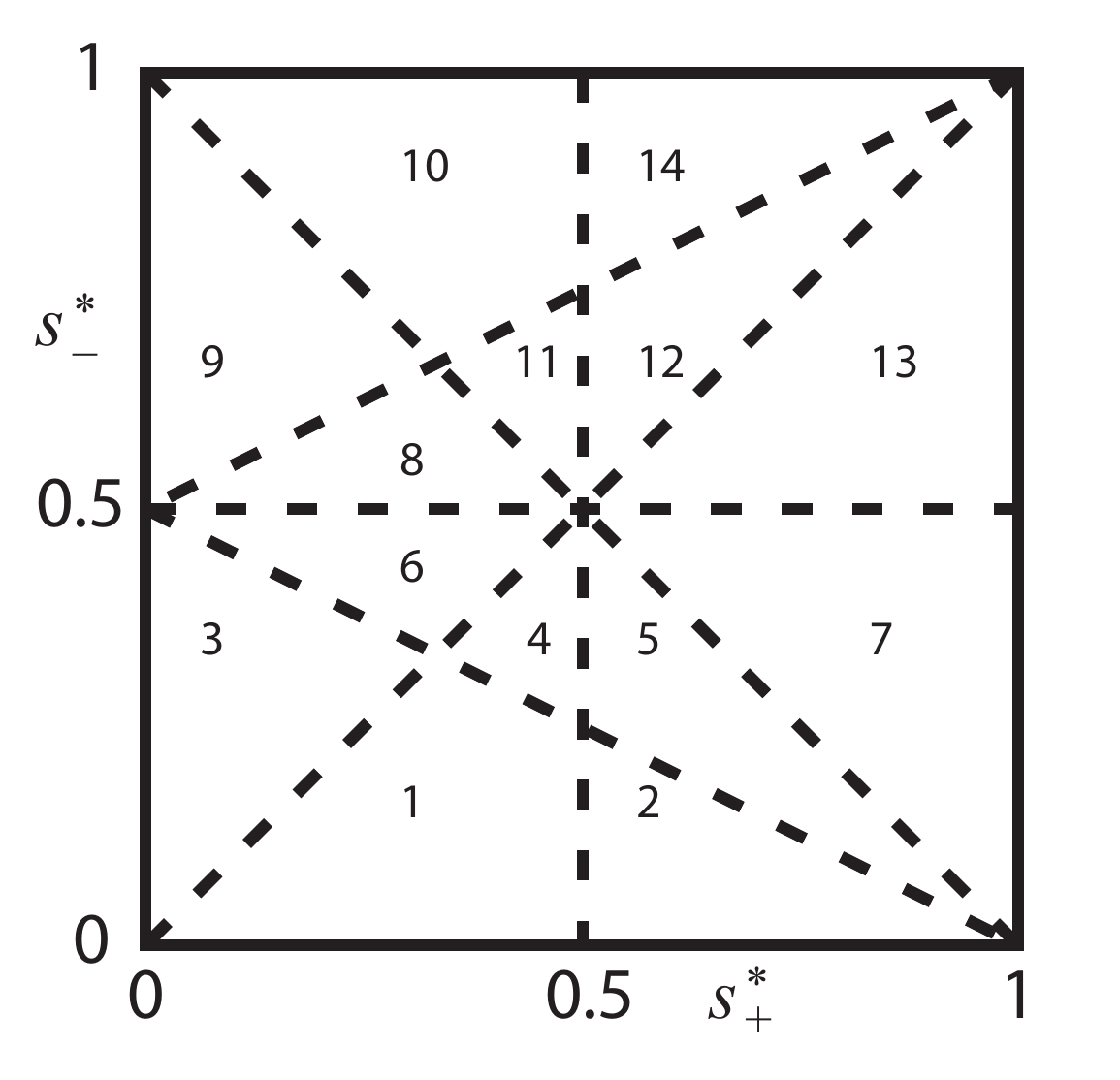}
\caption{The plane of parameters $0<s_+^* < 1$, $0 < s_-^* < 1$ split into subregions with different topology of $\Theta$-function conditions intersections, corresponding to different dynamic regimes. The ternary diagrams corresponding to each subregion can be seen in \fig{regions}.}
\label{diagram-14}
\end{figure}
\begin{figure*}
\centering
\includegraphics[width=17 cm]{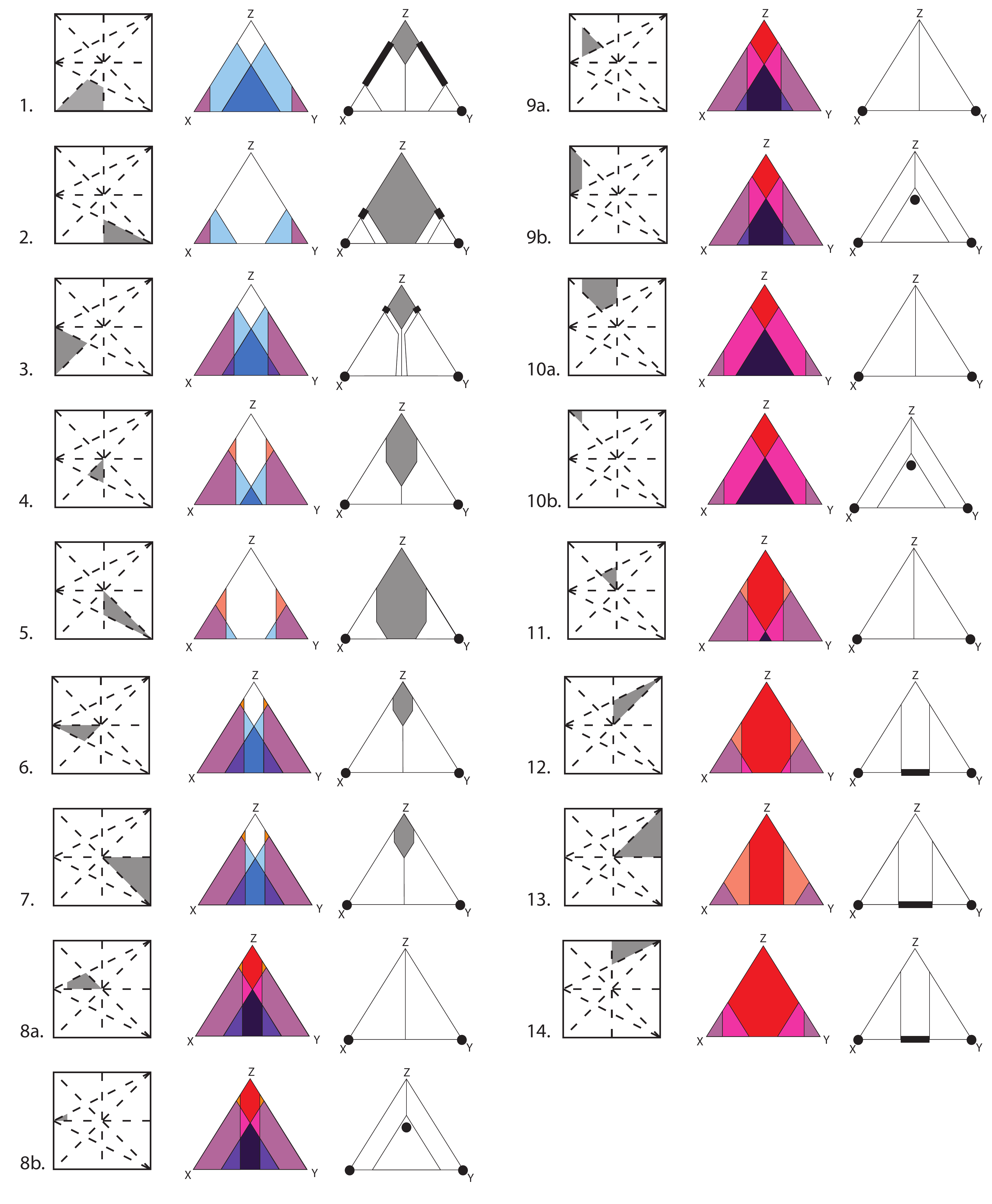}
\caption{List of topologically different ways of how the conditions under the $\Theta$-functions of Eqs.~\eqref{system_sym} may overlap and corresponding steady states and basins of attraction for each region of \fig{diagram-14}. 
First column: region on the $(s_+^*,s_-^*)$ plane, according to numbering in \fig{diagram-14}. 
Second column: the ternary X-Y-Z diagram with the position of the lines defined by the conditions in Eqs.~\eqref{system_sym}. The areas are color-coded according to corresponding pure states (see \fig{fig:pure}). Third column: attractors of the dynamics and their corresponding basins of attraction: black points represent attractive stable points; thick lines -- to stable points neutral in one direction (along the line) and attractive in the perpendicular one; gray areas -- to stable points neutral in both direction. Saddle points and other partially repulsive stable points are not shown in order not to overcomplicate the picture. Thin lines schematically delineate the borders of basins of attraction.}
\label{regions}
\end{figure*}

Second, if $x^*_+ + y^*_+>1$ the outcome additionally depends on whether the attractive point of regime 10, corresponding to Eqs.~\eqref{regime10}, is inside or outside the corresponding area (dark purple triangle) in \fig{fig:attr}. 
If it is outside, as shown in the second row of \fig{fig:attr}, then all trajectories starting in this region eventually end up on its boundary, switching to regime 7. 
As a result, in this case the only possible long-term outcome is the extinction of one of the languages (except for a saddle point at $x = x_+^*$, $y = y_+^*$, which is attractive when approached from regions of regimes 3 and 10 and repulsive when approached from regions or regime 7). 
Conversely, if the stable point is inside the triangular region, a stable multilingual equilibrium becomes possible. The condition for this to happen reads,
\begin{equation}
\left\{
\begin{aligned}
    & x_\infty = \displaystyle \frac{\gamma_x (1 - \kappa)}{\gamma_x (1 - \kappa) + \gamma_z \kappa + \kappa (1 - \kappa)} <x_+^*  \,  \\
    & y_\infty = \displaystyle \frac{\gamma_y \kappa}{\gamma_x (1 - \kappa) + \gamma_z \kappa + \kappa (1 - \kappa)}<y_+^* \, 
  \end{aligned}
\right. \, .
  \label{regime10cond}
\end{equation}
%
If $\gamma=\gamma_x=\gamma_y$ this reduces to
\begin{equation}
    \left\{
    \begin{aligned}
        & x_\infty = \displaystyle \frac{\gamma (1 - \kappa)}{\gamma+ \kappa (1 - \kappa)} <x_+^*  \, , \\
        & y_\infty = \displaystyle \frac{\gamma \kappa}{\gamma + \kappa (1 - \kappa)}<y_+^* \, .
      \end{aligned}
    \right. \, .
      \label{regime10cond2}
\end{equation}
%
It is easy to see that in this case all the trajectories, which start in regime 10, remain in this regime indefinitely and eventually get attracted to this stable point. Indeed, it suffices to notice that, according to Eqs.~\eqref{regime10}, the differences $x(t)-x_{\infty}$ and  $y(t)-y_{\infty}$ always decrease. 

Finally, consider a radically asymmetric situation, where attrition is always present but only for one of the languages (say, X). 
This corresponds to the combination of parameters $x_-^*=1$, $y_-^* = 0$ and might be possible, e.g., if Y (but not X), beyond providing means of communication within society, plays some significant external role like, for example, the language of knowledge or the language of some important external culture.

As summarized in Fig.~\ref{fig:asymattr}, the outcomes in this case are mostly controlled by the initial fraction $x_0$ of language X. 
In the case where $x^*_+ + y^*_+ > 1$ there are three possibilities. 
If $x_0 < 1-y_+^*$ (close to the YZ side of the ternary diagram), the dynamics converges to the monolingual stable point $y = 1$. 
If $x_0 > x_+^*$ (close to the X corner of the ternary diagram), it converges to the equilibrium of regime 5 [see Eqs.~\eqref{regime5}], which is an attractive stable point of the Y-Z subsystem. 
That is to say, here X speakers never turn bilingual, while Y speakers can. 
In the intermediate case, when $x_+^* > x_0 > 1-y_+^*$, all bilinguals initially present in the system, forget language X, resulting in the neutral equilibrium between monolinguals of language X and monolinguals of language Y. 
In the case when $x^*_+ + y^*_+ < 1$, only the first two of the three possibilities are present.

\begin{figure}
\centering
\includegraphics[width=7cm]{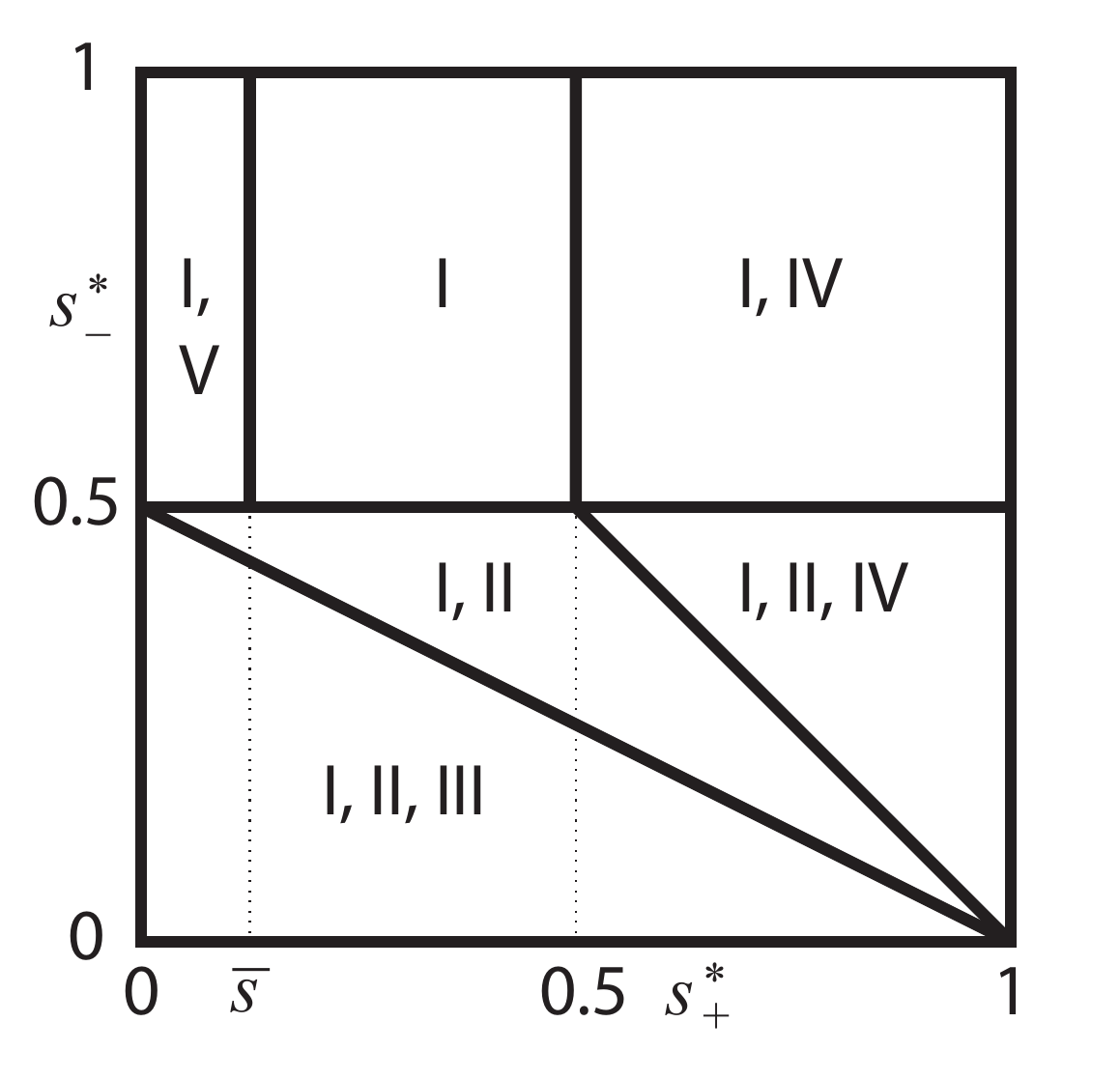}
\caption{State diagram of the model in the symmetric case. The plane of parameters $0<s_+^*<1$, $0<s_-^*<1$ is split into areas with different sets of possible long-term outcomes, depending on initial conditions. Outcome I is the extinction of one of the languages, outcome II -- frozen neutral equilibrium, outcome III -- coexistence of monolinguals of one language and bilinguals, outcome IV -- coexistence of two types of monolinguals with no bilinguals, outcome V -- attractive stable point with all three sub-populations surviving.  
}
\label{diagram}
\end{figure}
%

\subsection{Fully symmetric case} \label{sec:symmetric}

In the fully symmetric case when $\kappa = 1/2$, $\gamma_x = \gamma_y = \gamma$, $x_+^* = y_+^* = s_+$, $x_-^* = y_-^*=s_-$, the resulting symmetric equations \eqref{system_sym} have only three parameters, $s_{\pm}$ and $\gamma$, making the full classification traceable.
Depending on the specific values of $s_{\pm}$, the conditions under $\Theta$-functions in Eqs.~\eqref{system_sym} can split the X-Y-Z ternary diagram in 14 topologically different ways. 
The corresponding 14 regions of the $(s_{+},s_-)$ plane are shown in Fig.~\ref{diagram-14}. 
In Fig.~\ref{regions}, the first column outlines each of these regimes, the second column shows the corresponding ternary diagram and the third column shows the attractors and their basins of attraction (for brevity we skip the flow diagrams and do not show the unstable stationary points). 
Similarly to the constant attrition case discussed above, the outcome in three regions (8, 9 and 10) additionally depends on whether the stable point of pure regime 10 is outside or inside the corresponding domain; the corresponding two possibilities are shown in Fig.~\ref{regions} as options a and b, respectively. 
In the case of a fully symmetric system the conditions \eqref{regime10cond} separating these two regimes reduce simply to  
\begin{equation}
    s_+ > \displaystyle \bar{s} = \frac{2 \gamma }{4 \gamma + 1} \approx 2 \gamma \, ,
    \label{regime10sym}
\end{equation}
where the approximate equality corresponds to the $\gamma \ll 1$ limit.  
Note that the example considered in the previous section belongs to region 1.

Overall there are five possible attractors:

I. Full extinction of one of the languages; these are two attractive stable points at $(x,y,z) = (1,0,0)$ and $(x,y,z) = (0,1,0)$, marked by heavy dots at the X and Y corners of the ternary diagrams.

II. Frozen dynamics with all three concentrations fixed at their initial values; these stable points are fully neutral, the regions of these points are marked by a gray area in the diagrams.

III. Equilibrium between bilinguals and monolinguals of one of the languages with monolinguals of the other extinct language; these stable points are marked by thick lines on the XZ and YZ sides of the ternary diagrams; they are neutral in the direction of the line and attractive in the perpendicular direction.

IV. Community split: equilibrium between two types of monolinguals with bilinguals going extinct; these stable points are marked by thick lines on the XY side of the ternary diagrams; they are also neutrally stable in the direction of the line and attractive in the perpendicular direction.

V. Stable equilibrium between monolinguals of both types and bilinguals; this stable point is attractive, it is marked by a heavy dot inside the triangle. 

For each given combination of $s_+,s_-$ only some of these attractors  are observed. In Fig.~\ref{diagram} we show how the plane of parameters $(s_+,s_-)$ is split into regions with different sets of observed attractors.

Comparing these results to those of Ref.~\cite{Tamm-2025}, one sees that all three regimes observed in Ref.~\cite{Tamm-2025}, namely language extinction (I), frozen state (II) and stable language coexistence (V) are still possible in the three-state system. Moreover, there are two additional regimes. 
In one of them, bilinguals get completely extinct, returning to a two-state frozen regime (IV). 
In the other, speakers of the minority language become completely bilingual and, despite being able to communicate with each other in the majority language, continue to sustain the minority language indefinitely. 
Arguably, this is the most striking regime since it resembles the observed behavior in multiple bilingual societies where the fact that essentially the whole population is fluent in the majority language does not lead to extinction or visible decline of the minority one (see further discussion in \ref{sec:conclusion}. 

As mentioned above, full consideration of the general asymmetric case \eqref{system} is unfeasible. 
However, one might ask whether there are possible outcomes of the system dynamics which are different from the five described above for the fully symmetric case (clearly, they all are attainable at least if asymmetry is weak enough). Moreover, a neutral equilibrium between monolinguals of language Y and an equilibrated mix of bilinguals and monolinguals of language X (the attractor of pure regime 5) is also possible in highly asymmetric systems, see \fig{fig:asymattr} and discussion on it in Sec.~\ref{sec:limiting}. 
The remaining two possibilities (out of the 8 possible stable attractors for the pure states, see Sec.~\ref{sec:pure}) seem much more artificial. 
Indeed, the attractive point of regime 8, which implies that all population turns into bilinguals, is possible only if both learning thresholds are $x_+^* = y_+^* = 0$, i.e., both languages are still learned even if there are no monolinguals of these languages in the population, while the attractive stable point corresponding to the coexistence of monolinguals of one language and bilinguals (regime 9), requires one of those thresholds to be zero. 
That is to say, these attractive points, if possible at all, correspond to the very boundaries of the assumed parameter range.


\section{Discussion and Conclusion} \label{sec:conclusion}

%
\begin{table*}[ht!]
\centering
\scalebox{0.9}{
\begin{tabular}{||l | l |l | l | l | l||} 
\hline
\hline
Region& Sample& Minority  & Bilinguals & Majority  & Monolinguals\\
& size& monolinguals& & monolinguals& among minority\\
&&&&& language speakers\\
\hline
& $N$& $x$&$z$ &$y$ & $x/(x+z)$\\
\hline
Paraguay  &45744  &0.24\% & 8.45\% & 91.3\% & 0.028 \\
Galicia   &30850 & 3.59\% & 38.6\% & 57.8\% & 0.085 \\
Valencia  &60536  & 1.75\% & 14.2\% & 84.1\% & 0.110 \\
Basque Country  &22120  &3.04\% & 18.1\% & 78.9\% & 0.144 \\
Balearic Islands &13731   &4.99\% & 23.0\% & 72.0\% & 0.178 \\
Catalonia  &101688  &17.5\% & 37.5\% & 45.1\% & 0.318 \\
Quebec    &16848 &25.6\% & 33.6\% & 40.8\% & 0.433 \\
Finland   &15789 &1.31\% & 1.66\% & 97.0\% & 0.441 \\
Latvia    &15502 &13.4\% & 2.82\% & 83.8\% & 0.826 \\
Estonia   &2665 &25.5\% & 2.06\% & 72.5\% & 0.925 \\
Belgium   &41214 &32.9\% & 2.46\% & 64.6\% & 0.930 \\
Cyprus    &4227 &12.6 \% & 0.07 \% & 87.4 \% & 0.994 \\
\hline
\hline
\end{tabular}}
\caption{Prevalence of bilingualism in Twitter datasets collected in countries and regions with two dominant languages~\cite{Sanchez-2021}. The last column shows the fraction of minority language speakers who use only the minority language in their twits. Note that in 9 out of 12 examples one behavior (either monolingualism or bilingualism) dominates among the minority language speakers.}
\label{table:twitter}
\end{table*}

In this paper we generalized the threshold model of language competition introduced in Ref.~\cite{Tamm-2025} to the three-state model where the existence of bilinguals is taken into account explicitly. 
The model includes learning and attrition thresholds that encode the minimal exposure needed to acquire or maintain a language, and allows bilinguals to influence learning and attrition asymmetrically via different effective fractions of speakers for these two processes. 
The suggested threshold-based formulation ensures that the dynamics of language competition is described by a system of piecewise-linear differential equations, which are analytically tractable. Solution of the model consists of two stages.
First, we solve the system for each possible combination of values of $\Theta$-function in Eqs.~\eqref{system}, getting the solutions of what we call ``pure regimes''. At any given instance of time the actual solution of  Eqs.~\eqref{system} belongs to one of these regimes. Second, these solutions need to be glued together at points where variables $x$ and $y$ cross a threshold and the corresponding $\Theta$-function switches either on or off. As a result for any given set of parameters the full solution is a concatenation of the pure regime solutions.

Similarly to the conventional MW model, the suggested generalization contains a large number of independent parameters making full consideration rather cumbersome. However, it is possible to fully enumerate possible dynamic scenarios and long-term outcomes.
Qualitatively, our model can yield four broad classes of outcomes. 

The first one is language extinction, which is the only possible outcome in the conventional MW model, and is the most common one in our model as well (note, e.g., that for any set of parameters in the symmetric system the extinction is possible at least for some initial conditions). 

There are also two outcomes similar to those we predicted in the analogous two-state model \cite{Tamm-2025}: the completely frozen state, and the attractive stable point corresponding to stable language coexistence. 

Furthermore, there is a new class of mixed outcomes, corresponding to partially attractive and partially neutral stable points, i.e., stable points which form one-dimensional manifolds so that, while the outcomes are stable with respect to perturbation of the initial condition in one direction, perturbation in the other directions results in shifting the outcome along this manifold of stable points.

In the symmetric or weakly asymmetric systems there are two examples of such mixed outcomes. 
The first one (regime IV of the symmetric system) corresponds to a split of the society into two non-interacting monolingual communities. 
This state is attractive in the sense that bilinguals initially present in the system become extinct, and neutral in that the exact proportion in which the population is split between the two languages depends on the initial conditions. 
The second possibility (regime III of the symmetric system) is an equilibrium in which the monolingual sub-population of one language (say, X) remains monolingual, while all monolinguals of the language Y become bilingual. 
In this case, the equilibrium is stable with respect to changing the initial composition between monolinguals Y and bilinguals, but the exact split between monolingual X  and bilingual communities in the final state depends on initial conditions. 

This result is obtained under the assumption that there are ideally sharp $\Theta$-functional thresholds in the learning and attrition rates. 
If actual rates are more smeared S-shaped function of effective densities (see red lines in \fig{S-shape}) we expect, similarly to the two-state model, that the frozen state will degrade eventually but at a time scale that far exceeds the typical observational scales. 
In turn, intermediate states turn into long-living transient ones: first the system converges at a time scale of order at most $\gamma^{-1}$, then these states degrade towards attractive stable points (language extinction or coexistence) at much larger time scale controlled by a particular form of the S-shaped rates.  

Such long-lived intermediate states should be observed in reality: bilingual societies should disproportionally often be observed in the vicinity of one of the mixed stable points. 
That is to say, minority language communities should often be either predominantly functionally monolingual and thus separated from the majority (i.e., close to intermediate regime IV) or predominantly functionally bilingual (i.e., close to intermediate regime III), while situations where the minority language speakers are split into comparable number of bilinguals and monolinguals should be relatively rare. 
As we show in Table 1, which uses the data from Ref.~\cite{Sanchez-2021}, this indeed seems to be the case: out of 12 societies with two dominant languages, 3/4 demonstrate either clear domination of monolingualism among minority-language speakers (e.g., in Belgium or Estonia), or clear domination of bilingualism (e.g., in Galicia and Basque Country). Note that Twitter data is, clearly, a skewed sample of the population (e.g., probability of using Twitter might be language-dependent). However, we expect it to provide a good qualitative insight into dominance of monolingual and bilingual behavior within the minority community. 
Although a detailed comparison with the historical dynamics of language shift in these societies is impossible due to absence of long-term quality data, we believe that this qualitative observation is an important argument in support of the presence of intermediate semi-stable states predicted in this paper.

In the future, it would be interesting to study how thresholds in language learning and attrition might interplay with spatial inhomogeneities, demographic processes and network structure of society to give rise to further complexity in the dynamics of language shift, supporting data-driven calibration and policy counterfactuals, e.g., how changes in exposure or retention environments reshape the landscape of outcomes. 
Beyond language dynamics, the same thresholded three-state formalism should be applicable to other social and possibly biological systems where acquisition and maintenance are driven by exposure-dependent processes.

\begin{acknowledgments}
The authors acknowledge support from the Estonian Research Council through Grant PRG1059. We would also like to thank David S\'anchez	for directing us the data on bilingualism in Twitter communities. 
\end{acknowledgments}


\bibliography{references,references_2}
\bibliographystyle{unsrt}


\end{document}